\documentclass[aps,floatfix,showpacs]{revtex4}
\usepackage{natbib}
\usepackage{dcolumn}
\usepackage{graphicx}

\newcommand{\physrep}{Phys.~Rep.}

\newcommand{\be}{\begin{equation}}
\newcommand{\ee}{\end{equation}}
\newcommand{\bea}{\begin{eqnarray}}
\newcommand{\eea}{\end{eqnarray}}
\begin{document}
\title{A minimal set of invariants as a systematic approach to higher order gravity models: Physical and Cosmological Constraints}
\author{Jacob Moldenhauer\footnote{Electronic address: jam042100@utdallas.edu}, Mustapha Ishak\footnote{Electronic address: mishak@utdallas.edu}}
\affiliation{
Department of Physics, The University of Texas at Dallas, Richardson, TX 75083, USA}
\date{\today}
\begin{abstract}
We compare higher order gravity models to observational constraints from magnitude-redshift supernova data, distance to the last scattering surface of the CMB, and Baryon Acoustic Oscillations. We follow a recently proposed systematic approach to higher order gravity models based on minimal sets of curvature invariants, and select models that pass some physical acceptability conditions (free of ghost instabilities, real and positive propagation speeds, and free of separatrices). Models that satisfy these physical and observational constraints are found in this analysis and do provide fits to the data that are very close to those of the LCDM concordance model. However, we find that the limitation of the models considered here comes from the presence of superluminal mode propagations for the constrained parameter space of the models.
\end{abstract}
\maketitle
\section{introduction}
For the last decade, several cosmological observations \cite{observations} have been indicating that the expansion of the universe is accelerating. Cosmic acceleration constitutes one of the most challenging current problems in cosmology and all physics and several possible causes or explanations have been proposed, see for example the reviews \cite{reviews}. These include various dark energy components, modifications to general relativity at cosmological scales, or a true or apparent effect of inhomogeneities in the universe. 

In this paper, we consider higher-order gravity models (HOG) that have attracted much attention in the recent literature since they can have a late-time self-acceleration as an alternative to a dark energy component \cite{Biblio300,IshakMoldenhauer2008,Carroll2005,HOGgroupJAM,Faraoni2006b,Sotiriou2007}. The models are derived from curvature invariants that are more general than the Ricci scalar and can exhibit early-time inflation as well as late-time acceleration, e.g. \cite{Sotiriou2006,MengWang2004}. In these models, the dynamical equations are more elaborate than usual Friedmann equations and the acceleration is a consequence of a different coupling between matter and space-time curvature. Some early papers, see e.g. \cite{BirrellDavies1982,Stelle1977}, have discussed  the models within a more general context of unification theories for field quantization on curved backgrounds. From a point of view of phenomenology, some of the models were found to fit cosmological observations \cite{Mena2006,Shirata2005,Borowiec2006} and other models fit solar system tests \cite{NavarroVanAcoleyen2005} while other did not, e.g. \cite{Chiba2003}.

However, in view of the large number of possible models, it became clear that a systematic approach to the models was very desirable \cite{Faraoni2006b,Sotiriou2007}. Recently, the authors proposed in \cite{IshakMoldenhauer2008} a systematic approach based on minimal sets of invariants (bases) as the smallest independent building blocks to construct these models. Indeed, previous theorems from the theory of invariants in general relativity \cite{Debever1964,CarminatiMcLenaghan1991,ZakharyMcIntosh1997} showed that curvature invariants are not independent from each other and related by syzygies. For a given algebraic type of the Ricci tensor (e.g. the Segre classification \cite{Segre,Stephani2003}) and a given Petrov type of the Weyl tensor (i.e. symmetry classification of space-times), {\textit{e.g.}}\cite{Petrov1954,Pirani1957,Penrose1960,Stephani2003}, there exists a complete minimal independent set (basis) of these invariants in terms of which all the other invariants may be expressed. Interestingly, all the Friedmann-Lemaitre-Robertson-Walker space-times of interest are covered by the Petrov Type O and Segre Type A1-$[(1\,1\,1,\,1)]$ (see for example \cite{Stephani2003}) where the number of independent invariants reduces to two invariants.

In this paper, we use the systematic approach and select models free of ghosts, instabilities, and separatrices, and then compare the models to observations of supernova magnitude-redshift data, distance to the CMB surface, and Baryonic Acoustic Oscillations, including Hubble Key project and age constraints. 

\section{Higher-order gravity models based on minimal sets}

An approach for systematizing the study of HOG models was proposed by the authors, see \cite{IshakMoldenhauer2008}.  The method is based on a connection made between theorems in invariants theory in relativity and higher-order cosmological models.
We outline here the major ideas of this method and refer the reader to \cite{IshakMoldenhauer2008} for the detail.
The method involves the identification of the Petrov type corresponding to the symmetry of the spacetime, {\textit{e.g.}} \cite{Petrov1954,Pirani1957,Penrose1960,Stephani2003}, and the Segre type of the Ricci tensor \cite{Segre,Stephani2003}.  In short, the Ricci tensor and the Weyl tensor are considered as tensorial operators acting on space-time vectors and bi-vectors, respectively, where the types are determined from the multiplicity and properties of the eigenvalues and eigenvectors in each case, {\textit{e.g.}} \cite{Petrov1954,Pirani1957,Penrose1960,Stephani2003}. This limits the number of independent invariants that could be used to describe the spacetime.  This unique number of invariants can be used to build a basis for describing the spacetimes in higher-order gravity models and thereby allowing for the development of a systematic (taxonomy) method of constraining the otherwise infinite number of invariants possible to consider.  In fact, all Friedman-Lemaitre-Robertson-Walker (FLRW) manifolds fall under Petrov type O and Segre Type A1-$[(1\,1\,1,\,1)]$ with only two independent invariants, namely, $\{R,\,R1\}$ where 
\be
R=g^{\alpha\beta}R_{\alpha\beta},
\ee
is the Ricci scalar, and, the invariant
\be
R1=\frac{1}{4}S_{\alpha}\,^{\beta}\, S_{\beta}\,^{\alpha},
\ee
is built using the mixed trace-free Ricci tensor, $S_{\beta}\,^{\alpha}$.
  
Consistent with the basic idea, it was found in \cite{IshakMoldenhauer2008} that these independent invariants eliminate some unwanted redundancies in the generalized Friedmann equations and lead to compact formulations. 

 For illustration, let's give here the general field equations for $f(R,R1)$ models using the action of the form
\be
I=M_p\int{d^{4}x \sqrt{-g}[\frac{1}{2}R+f(R,R1)]+\int{d^{4}x}\sqrt{-g}(L_m+L_r)},
\label{eq:ActionRR1a}
\ee
where $M_p=(-1/8\pi G)$, we vary the action with respect to the metric, $g_{\alpha\beta}$, and we find the generalized field equations
\bea
\lefteqn{S^{\alpha\beta}-\frac{1}{4}g^{\alpha\beta}R-\frac{1}{2}g^{\alpha\beta}f+f_{R}S^{\alpha\beta}+\frac{1}{4}f_{R}g^{\alpha\beta}R+g^{\alpha\beta}f_{R;\gamma}\,^{\gamma}-f_{R;}\,^{\alpha\beta}+\frac{1}{2}f_{R1}S^{\alpha\gamma}S^{\beta}\,_{\gamma}+\frac{1}{8}f_{R1}S^{\alpha\beta}R}\nonumber\\
& &+\frac{1}{4}(f_{R1}S^{\alpha\beta})_{;\gamma}\,^{\gamma}+\frac{1}{4}g^{\alpha\beta}(f_{R1}S^{\gamma\delta})_{;\gamma\delta}-\frac{1}{4}(f_{R1}S^{\gamma\beta})_{;}\,^{\alpha}\,_{\gamma}-\frac{1}{4}(f_{R1}S^{\gamma\alpha})_{;}\,^{\beta}\,_{\gamma}=8\pi GT^{\alpha\beta},
\label{eq:FieldEquationsRR1a}
\eea
where we have used the definitions
\be
f_{R}\equiv\frac{\partial{f}}{\partial{R}},\,     f_{R1}\equiv\frac{\partial{f}}{\partial{R1}},
\ee 
and $T^{\alpha\beta}$ is the energy-momentum tensor.

In the field equations (\ref{eq:FieldEquationsRR1a}), the coupling between stress-energy and curvature is different than the one in GR (given by the first two terms on the LHS and energy momentum tensor on the RHS). The corresponding dynamics allows for some models to have late-time self-acceleration without the need for a cosmological constant or other forms of dark energy. In previous work \cite{IshakMoldenhauer2008}, the authors were able to show a range of these models in the basis $\{R,R1\}$ that showed self-acceleration at late times using numerical and analytical methods. The discussion there focused on power-law type solutions with late-time acceleration free from separatrix singularities, so that the models can transition from a matter dominated phase to the desired cosmic acceleration phase seen by current observations.  
In the minimal set basis $\{R,\, R1\}$, the authors of \cite{IshakMoldenhauer2008}, built many models in a systematic way by following previous examples in which the spin-2 ghost instability was avoided by construction. We follow up here on that and expand the discussion to other instabilities.  

\section{Physical constraints}
A ghost will refer to as a propagating degree of freedom which gives rise to a negative norm state upon quantization due to the wrong sign, leading to unphysical modes and particles \cite{Stelle1977,Chiba2005, DeFelice2006, Calcagni2006, NavarroVanAcoleyen2006, DeFelice2006b}. 
As it was discussed in, for example  \cite{DeFelice2006}, while a necessary condition to avoid ghosts is that the decoupled equations be free of fourth order derivatives, the remaining second order derivatives must also have the correct sign in order to have a true ghost free theory. In order to construct physically acceptable models, we restrict our models to the ones that comply with the two conditions and use some established results from previous studies \cite{Stelle1977,Chiba2005,Dvali2006,LiBarrowMota2007,NavarroVanAcoleyen2006,DeFelice2006,Uddin2009}. As usual, see for example \cite{Stelle1977,Chiba2005,Dvali2006,LiBarrowMota2007,NavarroVanAcoleyen2006,DeFelice2006,Uddin2009}, a theory with  an action built from Ricci scalar plus a function of another invariant, for example the Gauss-Bonnet (GB) invariant, i.e. $R+f(GB)$, can be re-written as the Einstein-Hilbert action plus the function coupled to a scalar field $\phi$ with potential $U(\phi)$, i.e. $R + f(\phi)\,GB\,-\,U(\phi)$. In the latter frame, the theory becomes like that of a Gauss-Bonnet one where the equations of motion (EOMs) then decouple into second order equations for the metric and for each scalar field involved. For example, \cite{DeFelice2006} derived some conditions for higher order invariants so that the EOMs do decouple into second order equations for the metric and each scalar field. It is worth clarifying that the decoupled equations are not the background field equations in the $R+f(GB)$ frame where the  generalized Friedmann equations contains higher order derivatives of the metric. For our present work, the conditions for ghost-free and non-superluminal propagations imposes some relations between the parameters that we are allowed to use in order to combine $R$ and $R1$. 

As discussed previously in \cite{IshakMoldenhauer2008}, the fourth-order derivatives in the decoupled equations and the corresponding unphysical states are avoided by construction if we choose functions of the Gauss-Bonnet-like form 
\be
a_1R^2+a_3(-\frac{5}{6}R^2-8R1),
\label{eq:GeneralGB1Form}
\ee
where we have imposed the same condition $a_2=-4a_3$ used in \cite{DeFelice2006} and where we used the vanishing of the Weyl tensor, see  \cite{IshakMoldenhauer2008}. Next, to build models that avoid the separatrix, we choose to set $a_1 \ne a_3$ which avoids the singularity at $\ddot{a}=0$ \cite{IshakMoldenhauer2008}. 
From now on, we limit the discussion to the case $a_1 \ne a_3=1$, and actions of the form
\be
f(R,R1)=\frac{m^{4n+2}}{[(a_1-\frac{5}{6})R^2-8R1]^n},
\label{eq:GeneralAction}
\ee
that are free of fourth order derivatives by construction.

Now, following the analysis of \cite{DeFelice2006}, we find, for our models, the equivalent equations to their equations (39), that are also necessary to have for a stable theory free of ghost, 
\be
1+\frac{2}{3}(6\beta-1)fR+8\ddot{f}>0,
\label{eq:GeneralGhostA}
\ee
and
\be
1+\frac{2}{3}(6\beta-1)fR+8H\dot{f}>0,
\label{eq:GeneralGhostB}
\ee 
where $H=\dot{a}/a$ is the Hubble parameter and a dot stands for derivative with respect to cosmic time.

Next, the spin-2 mode propagation speed is given by the ratio of the  coefficients of the  second derivatives, and the condition for a real positive and non-superluminal spin-2 propagation speed is thus given by 
\be
0 \le c_2^2=\frac{1+\frac{2}{3}(6\beta-1)fR+8\ddot{f}}{1+\frac{2}{3}(6\beta-1)fR+8H\dot{f}}\le 1.
\label{eq:GeneralSPIN2}
\ee

We also express, for our models, the condition for a real positive and non superluminal  condition for the scalar mode propagation speed \cite{DeFelice2006,HwangNoh2000,HwangNoh2005,Calcagni2006} that reads 
\be
0\le c_0^2=1+\frac{32\dot{f}\dot{H}}{3(\frac{2}{3}(6\beta-1)(\dot{f}R+f\dot{R})+8\dot{f}H^2)}-\frac{8(\ddot{f}-\dot{f}H)}{3(1+\frac{2}{3}(6\beta-1)fR+8H\dot{f})}.
\label{eq:GeneralSPIN0}
\ee

\begin{table*}[t]   
\begin{center}
\resizebox{7.0in}{!} {

\begin{tabular}{|c|c|c|c|c|c|c|c|}
\hline
Inverse HOG &Self-Acceleration & Deceleration During & Free  & Free of Ghost & Free of Ghost & Positive and real Spin-2 & Positive and real Spin-0
\\Models &at late times, &Matter Domination & of  &instability (\ref{eq:GeneralGhostA})& instability (\ref{eq:GeneralGhostB}) &propagation speed&propagation speed
\\for $n>0$ &$p>1$    & $1/2 \le p<1$&Separatrix & &   &  $c_2^2>0$&$c_0^2>0$   \\
\hline
\hline
$\beta=a_1-\frac{5}{6}$                   &&&&&& & \\ $n=1$&$a_1\le0.621$&$a_1\le0.621$&$a_1<\frac{5}{6}$& $ 0.065\le a_1 \le 0.621$& $0.467\le a_1 \le 0.621$& $0.467\le a_1 \le 0.621$&$ 0.563\le a_1 \le 0.621$\\ &&&&&&&\\
\hline
$\beta=a_1-\frac{5}{6}$ 
                    &&&&&&&\\ $n=2$&$a_1\le0.734$&$a_1\le0.734$&$a_1<\frac{5}{6}$& $ 0.474\le a_1 \le 0.734$& $0.636\le a_1 \le 0.734$& $0.636\le a_1 \le 0.734$&$ 0.683\le a_1 \le 0.734$\\ &&&&&&&\\
\hline
$\beta=a_1-\frac{5}{6}$
                    &&&&&&& \\ $n=10$&$a_1\le0.814$&$a_1\le0.814$&$a_1<\frac{5}{6}$& $ 0.745\le a_1 \le 0.814$& $0.776\le a_1 \le 0.814$& $0.776\le a_1 \le 0.814$&$ 0.788\le a_1 \le 0.814$\\ &&&&&&&\\
\hline

\end{tabular}
}
\end{center}
\caption{Physical parameter spaces for $f(R,R1)$ higher order gravity models (given by equation (\ref{eq:GeneralAction})). Conditions for self-acceleration, deceleration during matter domination, a cosmic evolution free of separatrices, free of ghost instabilities,  and requiring spin-2 and spin-0 modes to have propagation speed real and positive. 
The conditions are satisfied during the deceleration and acceleration phases of the cosmic evolution (we use the late-time accelerating solution in equation (\ref{eq:V_0GeneralSolution1}); for radiation-dominated and matter-dominated decelerating phases, we use $p=1/2$ and $p=2/3$, respectively). For each constraint indicated on the table, the allowed values for $a_1$ are given.}
\end{table*}

Now, we consider power-law solutions with $a(t)\propto t^p$ in flat FLRW spacetime for our accelerating branch as derived in our previous work \cite{IshakMoldenhauer2008} and as given further by the solutions (\ref{eq:V_0GeneralSolution1}). For radiation dominated and matter dominated decelerating phases, we use $p=1/2$ and $p=2/3$, respectively.

For the accelerating late time branch, we can write \cite{DeFelice2006},   
\be
c_2^2\approx \frac{(6\beta-1)p(2p-1)+8(n+1)(4n+3)}{p(6\beta-1)(2p-1)+8p(n+1)},
\label{eq:LTGeneralSPIN2}
\ee
and 
\be
c_0^2=1-\frac{32p(n+1)}{3[p(6\beta-1)(2p-1)(4n+2)+8p^2(n+1)]}-\frac{8(n+1)(4n+3-p)}{3[p(6\beta-1)(2p-1)+8p(n+1)]}.
\label{eq:LTGeneralSPIN0}
\ee

Finally, we recall here the discussion given in our previous work \cite{IshakMoldenhauer2008} for avoidance of separatrix singularities. The generalized Friedmann equations for actions of the form (\ref{eq:GeneralAction}) were shown in \cite{IshakMoldenhauer2008} to lead to the power law solutions with  
\be
p=\frac{1}{2}\pm\frac{\sqrt{6}}{12\sqrt{\beta}}, \frac{12\beta+3n+42\beta n+48\beta n^2\pm\sqrt{\Xi}}{24\beta(1+n)},
\label{eq:V_0GeneralSolution1}
\ee
\be
where \,\,\Xi=240\beta n+900\beta^2n^2+9n^2+588\beta n^2+480\beta n^3+2880\beta^2n^3+2304\beta^2n^4+24\beta. 
\ee

As discussed in \cite{IshakMoldenhauer2008}, in order to avoid singularities that lead to a separatrix, we must have no real solutions to the first two roots, (i.e. $\beta<0$ or equivalently $a_1<5/6$).  The remaining solutions, $a_1<5/6$, are divided into three regions depending on the type of attracting solutions. The first region gives negative attractors of no interest, and the second range has no real solutions, but the third range gives two real positive attractors with one large enough for self acceleration, see \cite{IshakMoldenhauer2008}. 

Our results are summarized in Table I and include conditions for imposing a decelerating phase followed by an accelerating one, a ghost-free evolution, a real and positive propagation speed for all modes, and separatrix-free models. And we find models that meet all these conditions during both accelerating and decelerating phases. However, also when both phases are included in the analysis, we find that the parameter space range for which all the modes propagate subluminally is in conflict with either the separatrix-free condition or the ghost-free ranges. In other words, we find that if we impose on the models to satisfy all the conditions above during both deceleration and acceleration phases, then there are modes with superluminal propagations within one phase of the other of the cosmic evolution.  

\section{Constraints From Supernova Magnitude-Redshift Data, Hubble Key Project, and Bounds on Age of the Universe}
One of the first compelling evidences for cosmic acceleration came from Supernovae type Ia observations.  We use the recent Union data set of supernovae \cite{Kowalski2008} to constrain the HOG models given in terms of the $\{R,R1\}$ basis.  We use the distance modulus as a function of the redshift $z$,
\be
\mu(z)=\textit{m}(z)-M=5\log_{10}{D_L(z)}+25,
\label{eq:DistanceModulus}
\ee
where $\textit{m}(z)$ is the magnitude of the supernova and $M$ is considered as a nuissance parameter degenerate with the Hubble parameter, $H_0$, and also with our parameter $\hat{m}$ defined below, $D_L(z)$ is the luminosity distance in units of Mpc calculated for the HOG models and given by
\be
D_L(z)=(1+z)\int^z_0 \frac{1}{H(z')}dz'.
\label{eq:LuminosityDistance}
\ee
where $H(z')$ is the solution to the non-linear differential generalized Friedmann equation for HOG models (see for example Eq. (\ref{eq:GeneralFriedmanna}) below), similar to the Friedmann equation for GR models.  We restrict ourselves to studying models that fit the physical constraints discussed earlier in this paper, and in earlier work \cite{IshakMoldenhauer2008}. In view of the complexity of the analysis, we focus here on the cases of $n=1$ and $n=2$ in these models. 

For $n=1$, the generalized Friedmann equation reads (again, we use the notation $\beta \equiv a_1-5/6$),  
\bea
&&3H^2-\frac{m^{6}}{6(6\beta\dot{H}^2+24\beta H^2\dot{H}+24\beta H^4-\dot{H}^2)^{3}}\Big{(}6048\beta^2H^6\dot{H}+1152\beta^2H^8-240\beta H^2\dot{H}^3-360\beta H^4\dot{H}^2\nonumber\\
&&-6H^2\dot{H}^3+5616\beta^2H^4\dot{H}^2+3\dot{H}^4+864\beta^2H^5\ddot{H}+6H\dot{H}^2\ddot{H}+1656\beta^2H^2\dot{H}^3-144\beta H^3\dot{H}\ddot{H}+216\beta^2H\dot{H}^2\ddot{H}\nonumber\\
&&-72\beta H\dot{H}^2\ddot{H}+864\beta^2H^3\dot{H}\ddot{H}-36\beta \dot{H}^4+108\beta^2\dot{H}^4+48\beta H^5\ddot{H}+144\beta H^6\dot{H}\Big{)}=8\pi G\rho_m +8\pi G\rho_r.
\label{eq:GeneralFriedmanna}
\eea

We use in the analysis a modified version of the Markov Chain Monte Carlo (MCMC) package CosmoMC \cite{LewisBridle2001}. We customized this package for our HOG models and use it to constrain their parameters.  We found that an elaborate and essential step in our analysis was to numerically integrate the stiff ODE (\ref{eq:GeneralFriedmanna}) in order to derive the Hubble expansion rate for a wide range of redshift. For this task, we followed the same  notation and parameterizations as Ref. \cite{Mena2006}, making a necessary change of variable in our integration and also using a good approximation at higher redshifts in order to obtain initial values for our codes to perform stable integrations from $z=5$ down to $z=0$. 
\begin{figure}
\begin{center}
\begin{tabular}{|c|c|}
\hline

{\includegraphics[width=2.8in,height=2.8in,angle=0.]{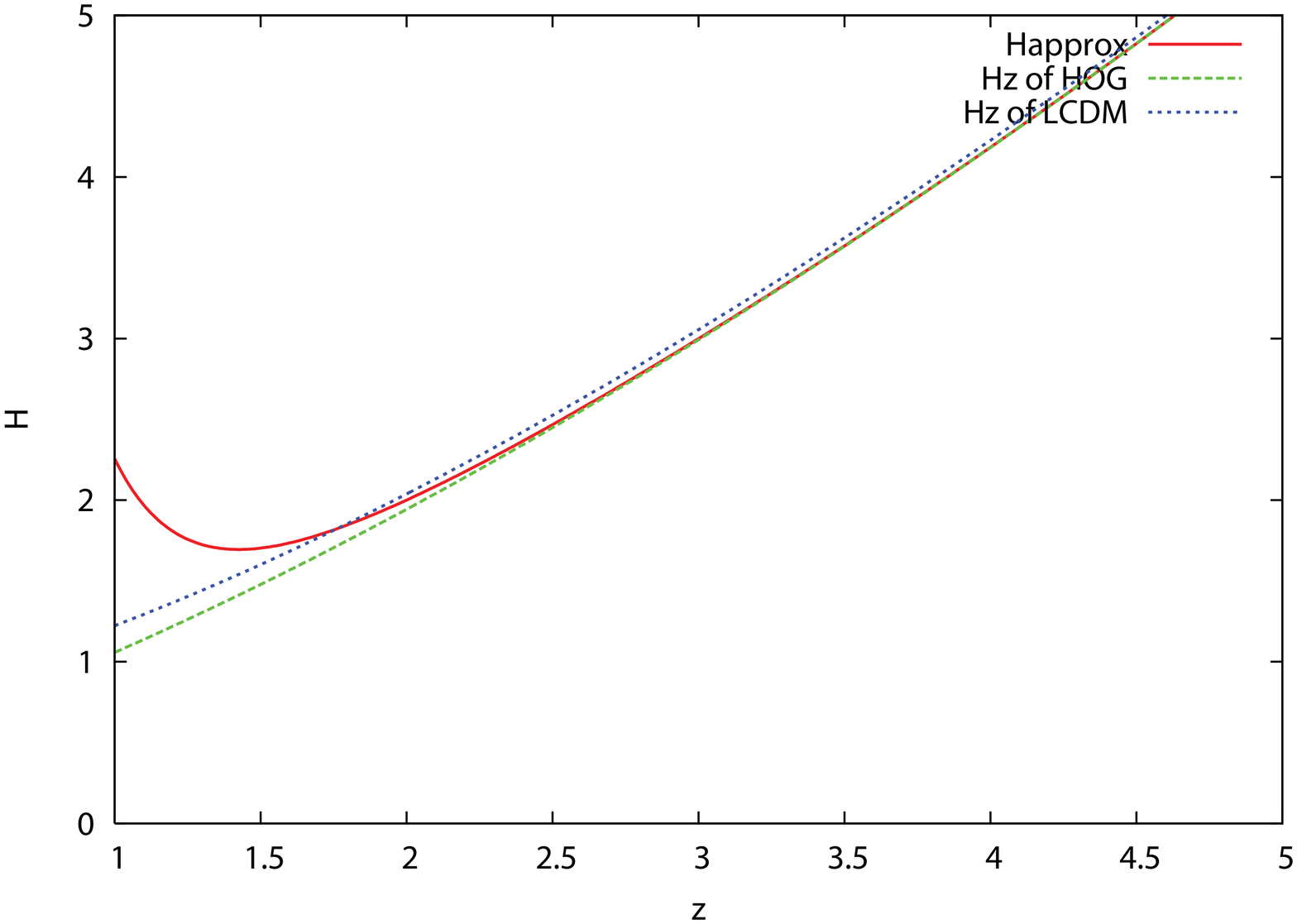}} &
{\includegraphics[width=2.8in,height=2.8in,angle=0.]{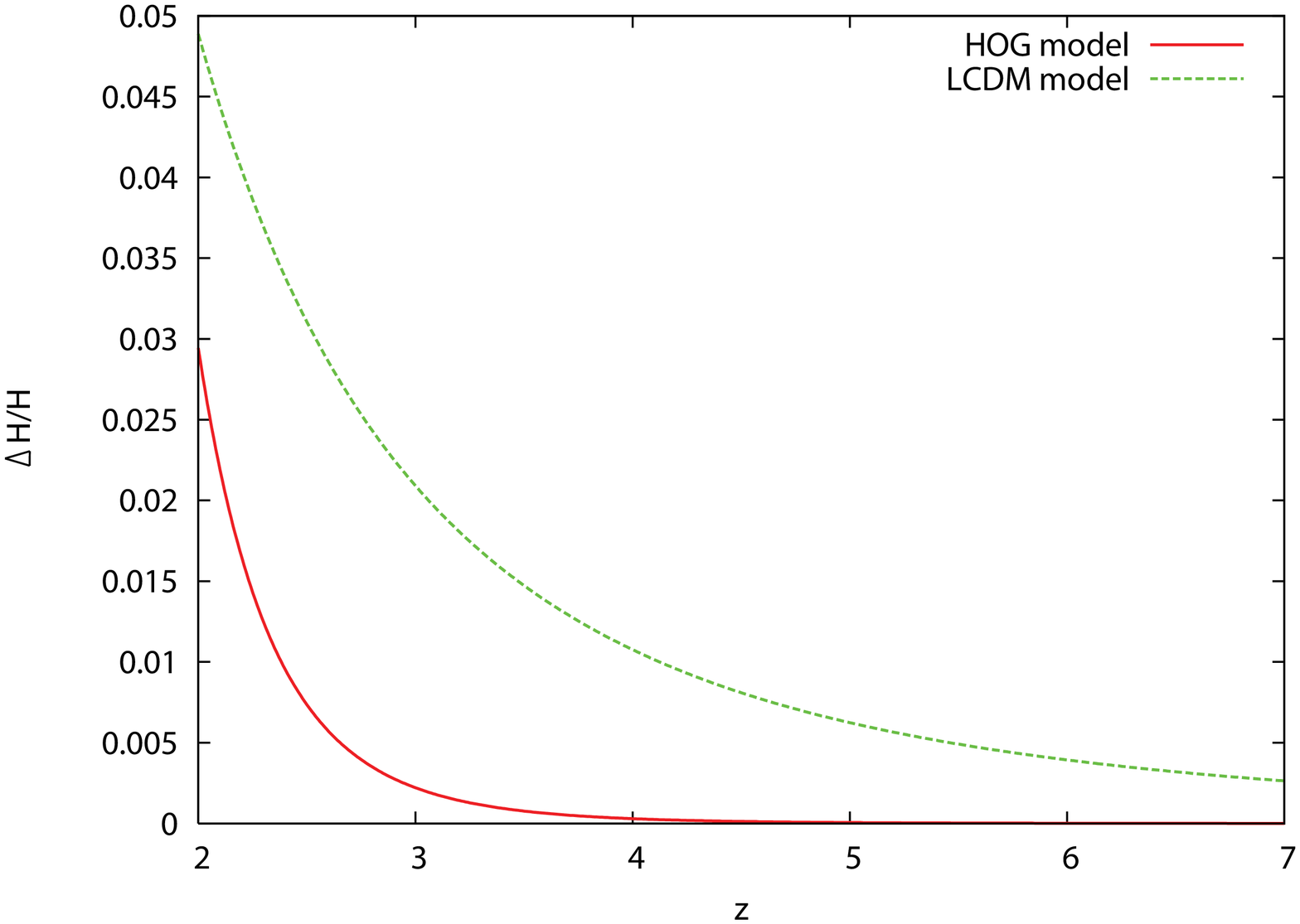}} \\
\hline
\end{tabular}
\caption{
LEFT: This figure shows how the approximation $H_{approx}$ of equation (\ref{eq:Happrox}) fits well the integrated $H(z)$ from the ODE for our HOG models for redshifts $>5$. RIGHT: The relative difference $\frac{H_{approx}-H(z)}{H(z)}$ shows that for $z>5$  the fit is better than $0.001$.  We use this approximation at high redshifts and a full numerical integration for $z\le5$. The relative difference between LCDM and HOG models is small at high redshift, and becomes important at late times. } 
\end{center}
\end{figure}
Indeed, in order to deal with the stiffness of the problem, we replaced $H$ in (\ref{eq:GeneralFriedmanna}) by the logarithmic variable $u=\ln{(H/\hat{m})}$ (see \cite{Mena2006}) where  
\be
\hat{m}=m/(12a_1-10)^{1/6}. 
\label{eq:mhata1}
\ee
Further, following \cite{Mena2006}, we define
\be
\alpha=\frac{12a_1-12}{12a_1-10},
\label{eq:alphaa1}
\ee
and
\be
\sigma=sgn(12a_1-10),
\label{eq:sigmaa1}
\ee
which give the relationship for our parameter $a_1$ as 
\be
a_1=\frac{5\alpha-6}{6(\alpha-1)}.
\label{eq:a1alphasigma}
\ee
This allows one to write the sources term, see \cite{Mena2006}, as  
\be
\tilde{u}=\ln{(\tilde{\omega}_r e^{-4N} + \tilde{\omega}_m e^{-3N})}/2,
\ee
where $N=\ln{a}$ and 
\be
\tilde{\omega}_m\equiv \frac{8\pi G}{3}\frac{\rho_0}{\hat{m}^2}.
\label{eq:wbarsource}
\ee
Further, with $\omega_m = \Omega_m h^2$ and $h=H_0/(100 km/s/Mpc)$, one writes 
\be
\tilde{\omega}_m=\frac{\omega_m}{\hat{m}^2}.
\label{eq:wbar}
\ee
Similarly, $\tilde{\omega}_r$ is defined for radiation but we consider its contribution to be negligible at late times.

\begin{figure}
\begin{center}
\begin{tabular}{|c|c|c|}
\hline

{\includegraphics[width=2.0in,height=2.0in,angle=0]{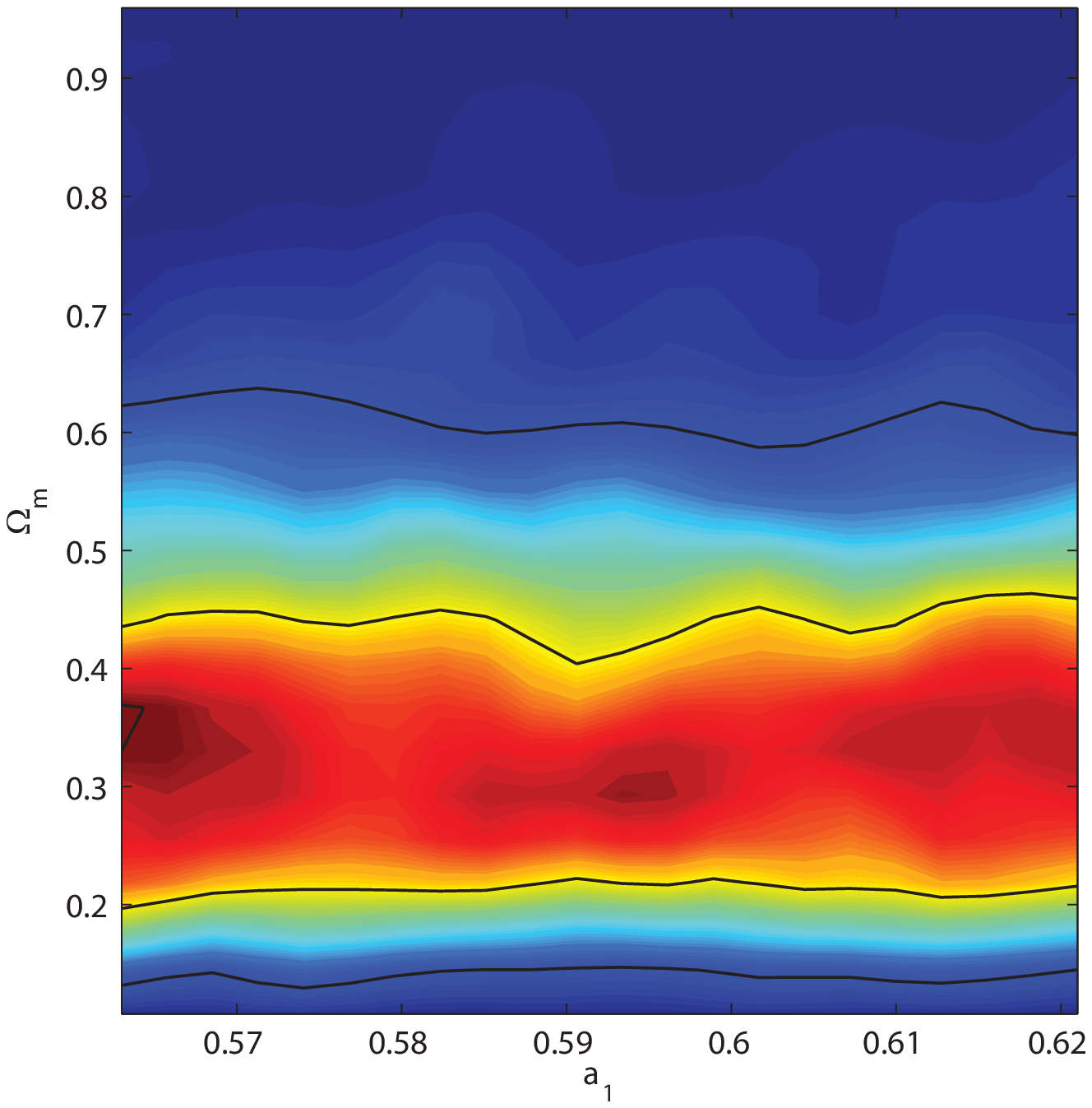}} &
{\includegraphics[width=2.0in,height=2.0in,angle=0]{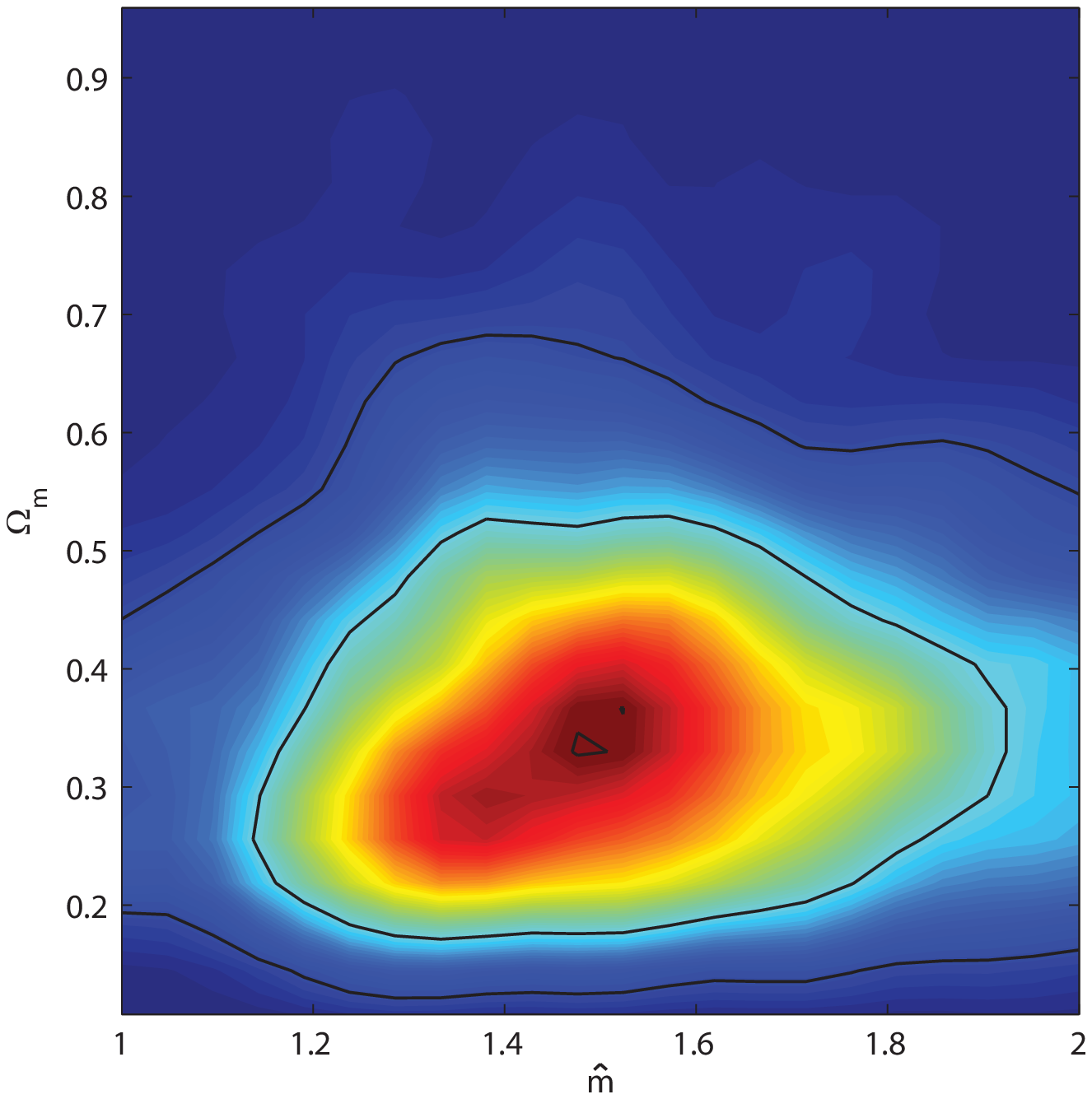}} &
{\includegraphics[width=2.0in,height=2.0in,angle=0]{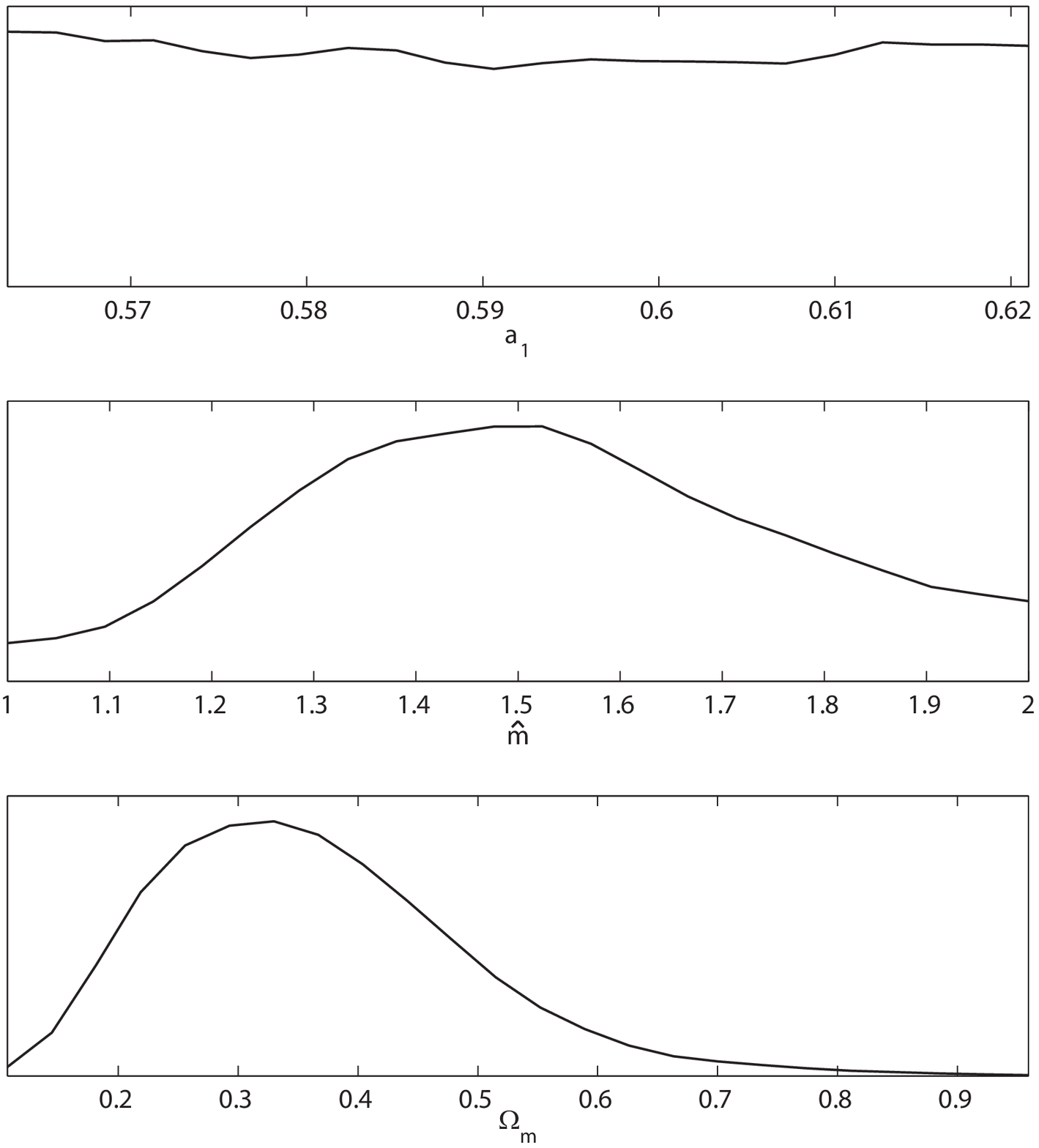}} \\
\hline
\end{tabular}
\caption{LEFT: HOG models with $n=1$. $2D$ joint contour plots for ($a_1$,$\Omega_m$) plane for, Union SNe Ia data sets, Hubble Key Project, and age of the universe where the inner and outer loops are $68\%$ and $95\%$, respectively.  CENTER: Same as LEFT, but for the ($\hat{m}$,$\Omega_m$) parameter space. RIGHT: 1D parameter distributions for the same data set as at LEFT and CENTER.} 

\end{center}
\end{figure}

Using this change of variable and the notation above, the stiff ODE (\ref{eq:GeneralFriedmanna}) can be re-written as \cite{Mena2006}   
\be
u''y_1(u')+y_2(u')+18\sigma(y_3(u'))^3 e^{6u}(e^{2(\tilde{u}-u)}-1)=0, 
\label{eq:MenaUdiffeq}
\ee
where 
\bea
&&y_1(x)= 8\Big(27(a_1-1)^2 x^2+18(6a1-5)(a1-1)x+ 2(9a1-7)(6a1-5)  \Big)\Big/(6a_1-5)^2\\
&&y_2(x)=4\Big(135(a1-1)^2 x^4+ 6(141 a1-116)(a1-1)x^3+ 2(6a1-5)(153a1-133)x^2\nonumber\\
&&+12(21a1-17)(6a1-5)x+8(6a1-5)^2\Big)\Big/(6a_1-5)^2\\
&&y_3(x)= 2\Big(3(a_1-1)x^2+2(6a_1-5)(x+2)\Big)\Big/(6a_1-5),
\label{eq:yxDiff}
\eea
and $'=d/dN$.  The final forms (\ref{eq:MenaUdiffeq})-(\ref{eq:yxDiff}) above are similar to what was used in \cite{Mena2006} but expressed in terms of our parameters in the basis $\{R1,R\}$. As mentioned earlier, unlike for equation (\ref{eq:GeneralFriedmanna}), we found that the form above allows a stable numerical integration form redshifts of $z=5$ down to $z=0$. We also found it necessary to perform the integration forward in time (backward in the redshift) with initial  conditions provided by the approximate solution given in \cite{Mena2006} and expressed in our notation as 
\be
H_{approx} = \hat{m}e^{\tilde{u}} \Big{(}1 + \frac{e^{-6 \tilde{u}}\tilde{u}''y_1(\tilde{u}')+y_2(\tilde{u}')}{36\sigma\,\, (y_3(\tilde{u}'))^3}\Big{)}.
\label{eq:Happrox}
\ee
As shown in Fig. 1, we verified that at higher redshifts ($z>2-3$) the approximate solution provides an excellent fit to the numerical solution of the ODE (\ref{eq:MenaUdiffeq}). For our models, we thus use (\ref{eq:Happrox}) in order to find the initial conditions for $u$ and $u'$ at $z=5$ and then start a numerical integration of (\ref{eq:MenaUdiffeq}) down to the supernova locations. Proceeding in this way, we obtained very stable programs to derive various Hubble plots for the HOG models. 

In a similar way, we derive the equations for the HOG models with $n=2$.
The modified Friedmann equations read
\be
u''y_4(u')+y_5(u')+\frac{432}{(6a_1-5)}\sigma(y_6(u'))^4 e^{10u}(e^{2(\tilde{u}-u)}-1)=0,
\label{eq:MenaUdiffeqn2}
\ee
where 
\bea
&&y_4(x)= 12\Big(15(a_1-1)^2 x^2+10(6a1-5)(a1-1)x+ 2(5a1-4)(6a1-5)  \Big)\\
&&y_5(x)=3\Big(135(a1-1)^2 x^4+ 36(23 a1-19)(a1-1)x^3+ 36(6a1-5)(8a1-7)x^2\nonumber\\
&&+8(27a1-22)(6a1-5)x+4(6a1-5)^2\Big)\\
&&y_6(x)= 3(a_1-1)x^2+2(6a_1-5)(x+1)\Big).
\label{eq:yxDiffn2}
\eea
And the expression for the approximation is given by  
\be
H_{approx} = \hat{m}e^{\tilde{u}} \Big{(}1 + (6a_1-5)\frac{e^{-10 \tilde{u}}\tilde{u}''y_4(\tilde{u}')+y_5(\tilde{u}')}{864\sigma\,\, (y_6(\tilde{u}'))^4}\Big{)}.
\label{eq:Happroxn2}
\ee

We start here with Supernova data combined with the Hubble parameter value $H_0=72\pm8 km/s/Mpc$ from Hubble Key Project \cite{HubbleKeyProject} and the prior on the age of the universe so that only models with $10<t_0<20 Gyr$ are considered. 
We use the Union set of supernovae which was compiled to gather the best supernovae from different surveys, including Supernovae Legacy Survey, ESSENCE Survey, HST, and other older sets \cite{Kowalski2008}.  After selection cuts the 414 SNe Ia are reduced to 307. As usual, the fitting of the SNe Ia uses the standard $\chi^2$ given by  
\be
\chi_{SN}^2=\sum^{i=307}_{i=1} \frac{[\mu^i_{HOG}(z)-\mu^i_{obs}(z)]^2}{\sigma^2}
\label{eq:ChiSquared}
\ee
where $\sigma$ is the magnitude uncertainty and $i$ the number of data points compared. 
\begin{figure}
\begin{center}
\begin{tabular}{|c|c|c|}
\hline

{\includegraphics[width=2.0in,height=2.0in,angle=0]{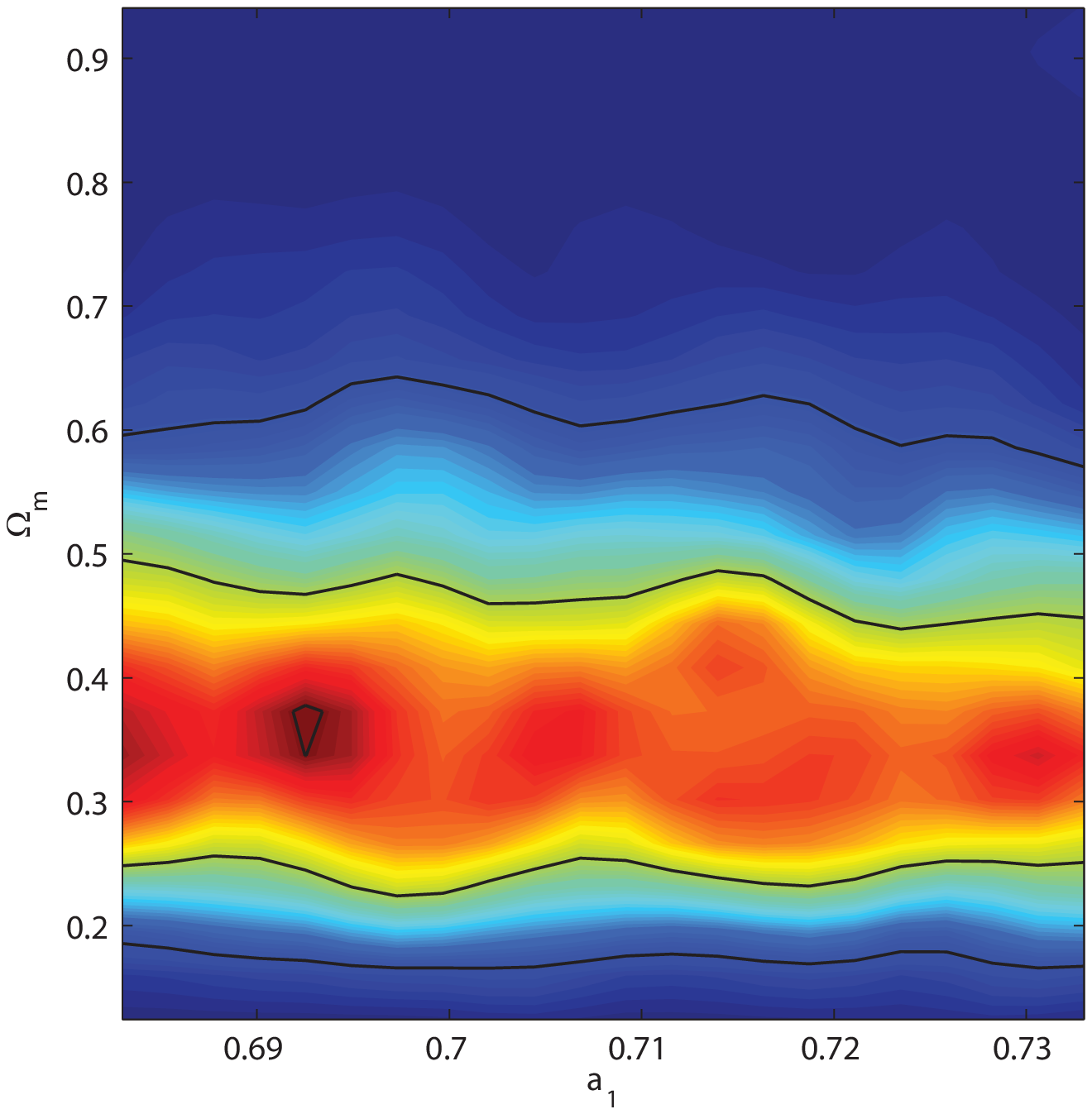}} &
{\includegraphics[width=2.0in,height=2.0in,angle=0]{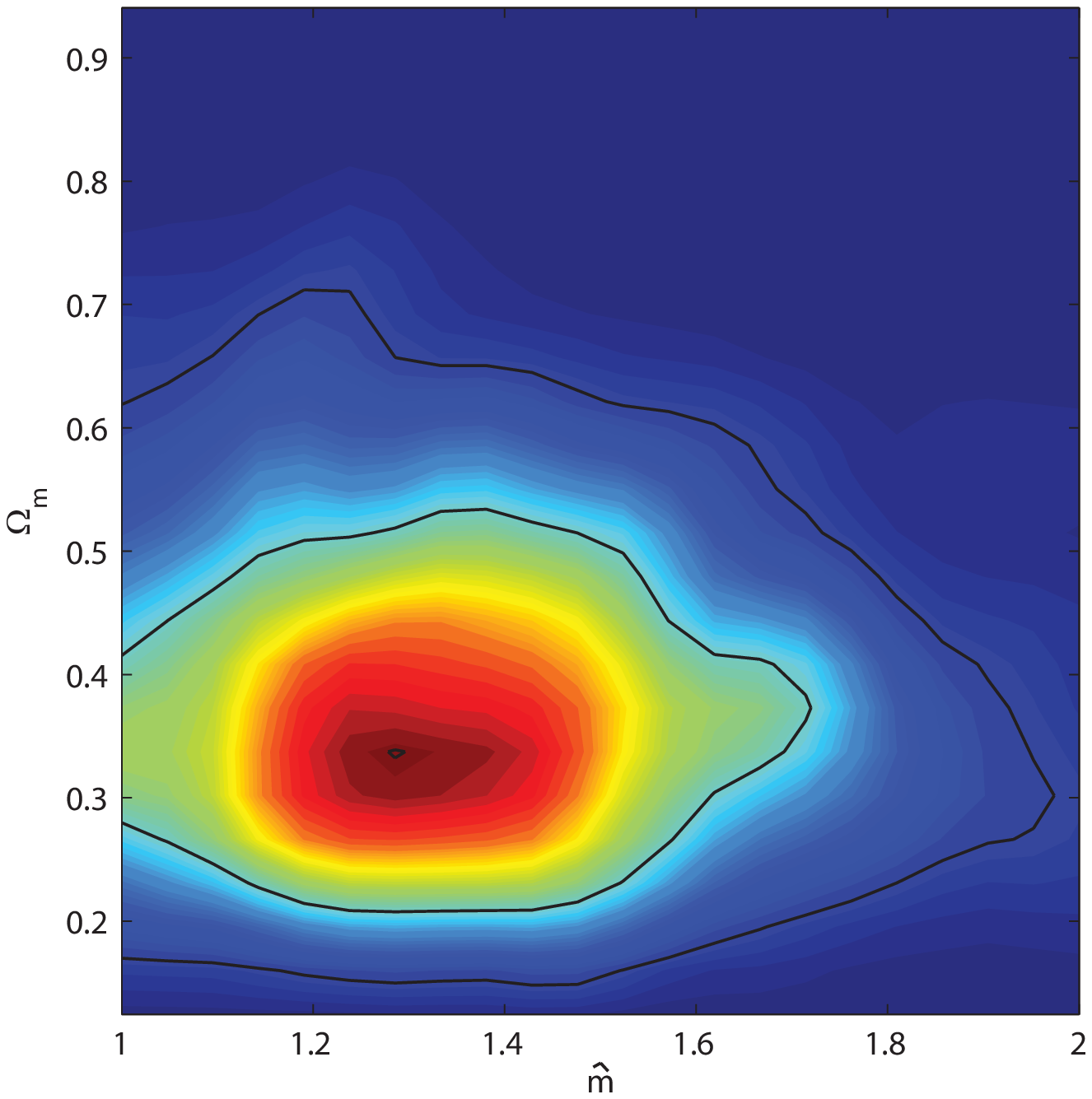}} &
{\includegraphics[width=2.0in,height=2.0in,angle=0]{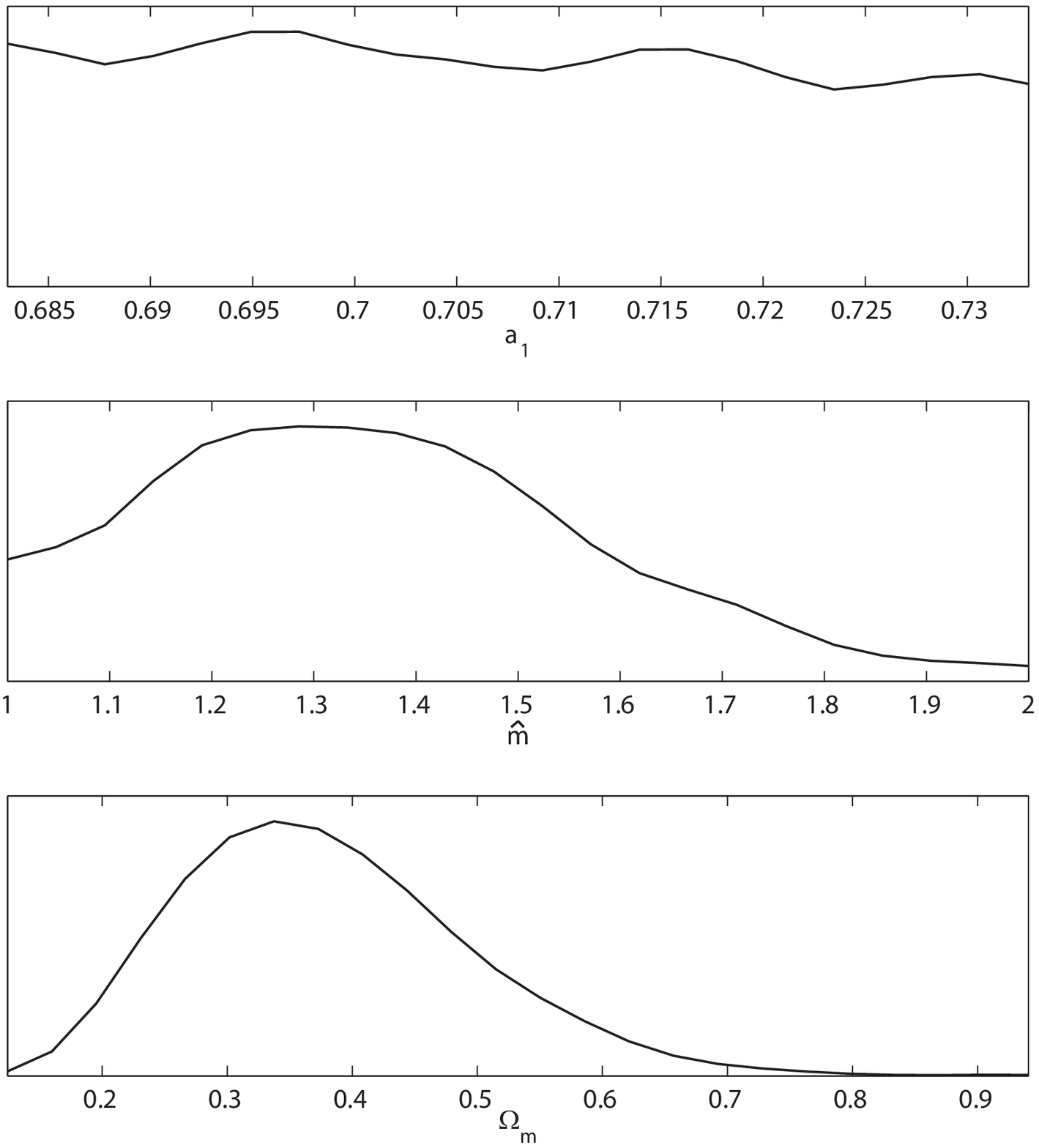}} \\
\hline
\end{tabular}
\caption{LEFT: HOG models with $n=2$. $2D$ joint contour plots for ($a_1$,$\Omega_m$) plane for, Union SNe Ia data sets, Hubble Key Project, and age of the universe where the inner and outer loops are $68\%$ and $95\%$, respectively.  CENTER: Same as LEFT, but for the ($\hat{m}$,$\Omega_m$) parameter space. RIGHT: 1D parameter distributions for the same data set as at LEFT and CENTER.} 

\end{center}
\end{figure}

Using the probes above, we use our modified version of CosmoMC to derive constraints on $\Omega_m$, $\hat{m}$ and $a_1$ for the HOG models with $n=1$ and $n=2$. 

For $n=1$, we restrict the model parameter $a_1$ to the region that passes the physical acceptability conditions given in Table I (see the last column). We find a best fit value for $\Omega_m=0.396^{+0.255}_{-0.242}$, in agreement with other studies using HOG models \cite{KoivistoMota2007a,KoivistoMota2007b}, but in need of tighter constraints as we do in the next section. The best-fit value $\hat{m}=1.555^{+0.429}_{-0.464}$ is somewhat well constrained in the parameter space shown in Fig. 2, and within $1\sigma$ deviations of twice the unitless Hubble constant $h=H_0/100 km/s/Mpc$, in consistency with \cite{Mena2006} for other models. While the fits do return a best value of $a_1=0.575^{+0.046}_{-0.012}$, we see in Fig. 2, the parameter $a_1$ is not well constrained for this cosmological data. In fact, we find that all the physical parameter space for $a_1$ fits well to the data. We find a best-fit $\chi_{SN}^2\approx311.7$ for our HOG model with fitting values given in Fig. 2. We procceed in a similar way for $n=2$ models, and find a best fit value $\Omega_m =0.239^{+0.434}_{-0.062}$ (consistent with the one above within the error bars) and the model parameter $\hat{m}=1. 217^{+0.662}_{-0.198}$ is about $\sim 2h$ again. The constraints are shown in Fig. 3. We find the best fit value for the model parameter $a_1=0.689^{+0.044}_{-0.006}$, but again, is not well constrained by the data presented in this section.  All the physical parameter space allowed by the constraints in Table I fits well to the data with a $\chi_{SN}^2\approx311.4$. The $\chi_{SN}^2$ for $n=1$ and $n=2$ models are close to the $\chi_{SN}^2\approx308.1$ obtained by the LCDM concordance model. In view of the of the possible systematic uncertainties in the supernova data, it is not clear that the difference between the two $\chi^2$ is significant. Our Hubble plot for the best fit HOG models with $n=1,2$ is presented in Fig. 4. 
\begin{figure}
\begin{center}
\begin{tabular}{|c|c|}
\hline

{\includegraphics[width=4.0in,height=4.0in,angle=0.]{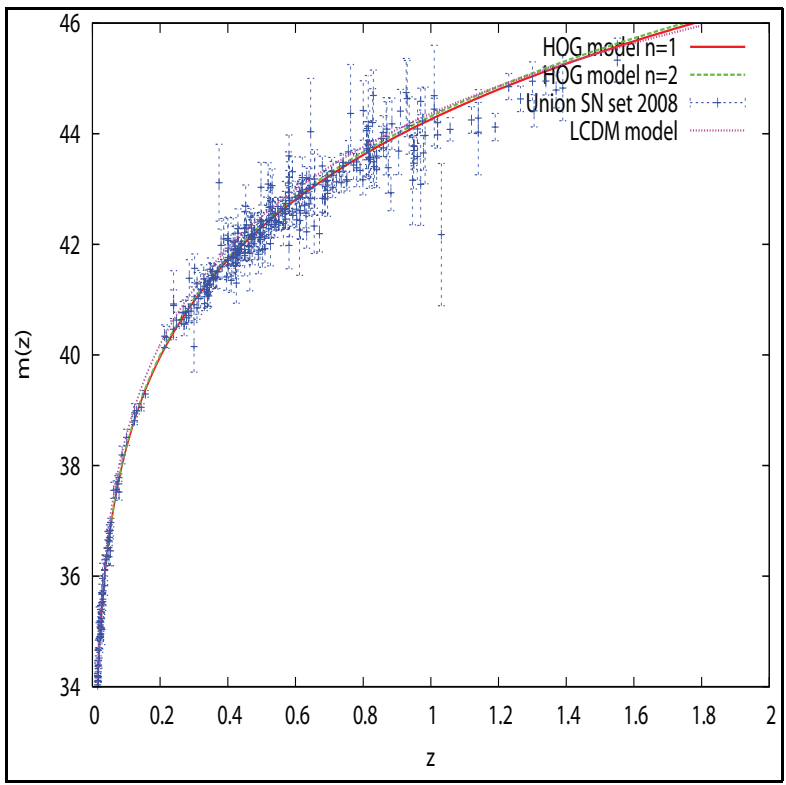}} \\
\hline
\end{tabular}
\caption{Hubble plot of best-fit cosmological parameters from combinations of Union SNe Ia data sets, Hubble Key Project, and age of the universe with best fit $a_1$ values for $n=1,2$ and comparisons to supernovae data with error bars and current LCDM model.} 
\end{center}
\end{figure}

\section{Adding constraints from distance to the CMB last scattering surface and Baryon Acoustic Oscillations}

In addition to supernova data constraints, Hubble Key project and age bounds, we consider here constraints from the distance to the CMB surface of last scattering and also Baryon Acoustic Oscillations (BAO). 
To do this, we use again a modified version of CosmoMC where various distance determinations are done using our numerical integrators as discussed above. We combine the approximation (\ref{eq:Happrox}) at high redshift to the numerical integration of (\ref{eq:MenaUdiffeq}) at lower redshifts down to zero. 

\begin{figure}
\begin{center}
\begin{tabular}{|c|c|}
\hline
{\includegraphics[width=2.8in,height=2.8in,angle=0]{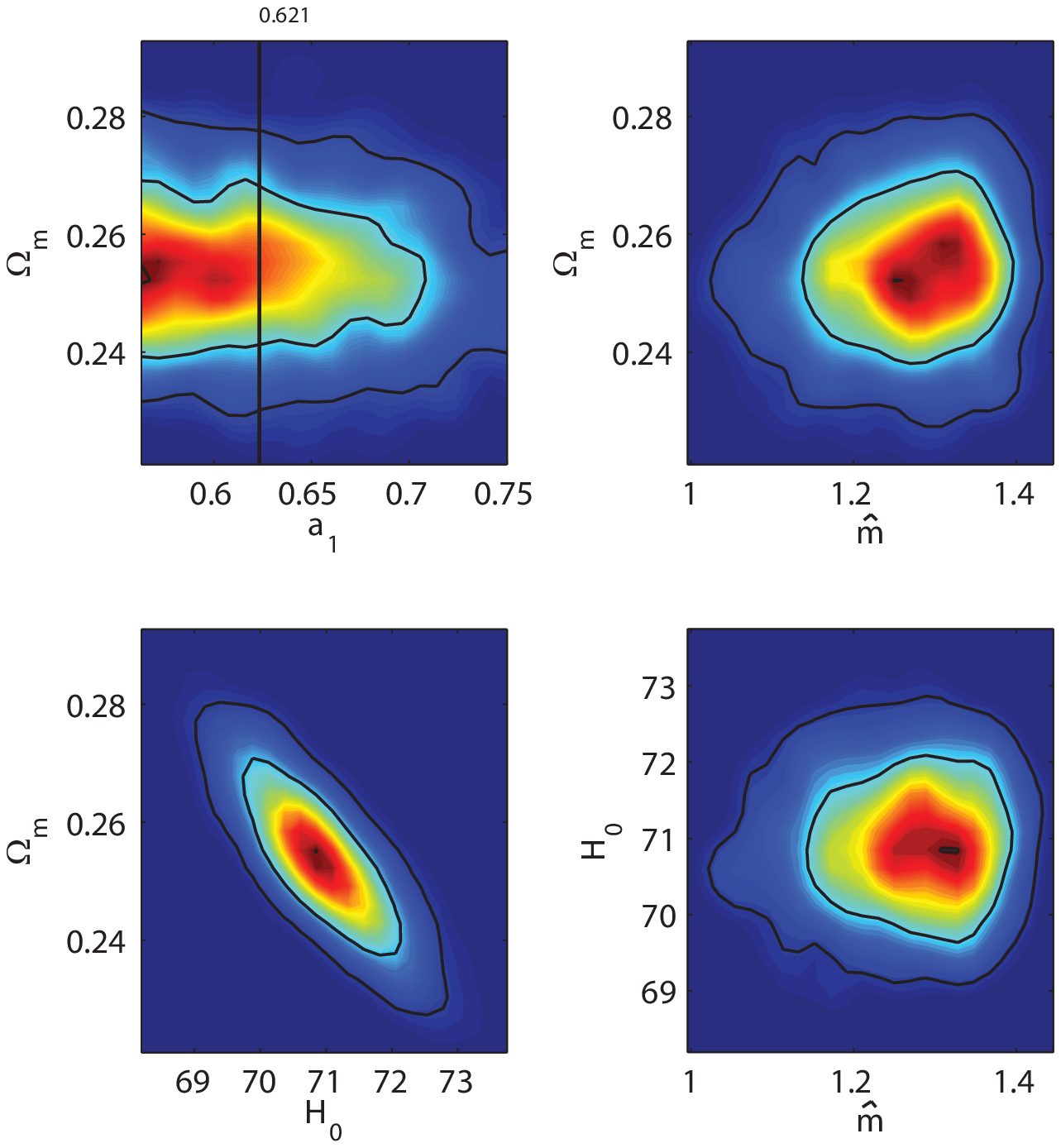}} &
{\includegraphics[width=2.8in,height=2.8in,angle=0]{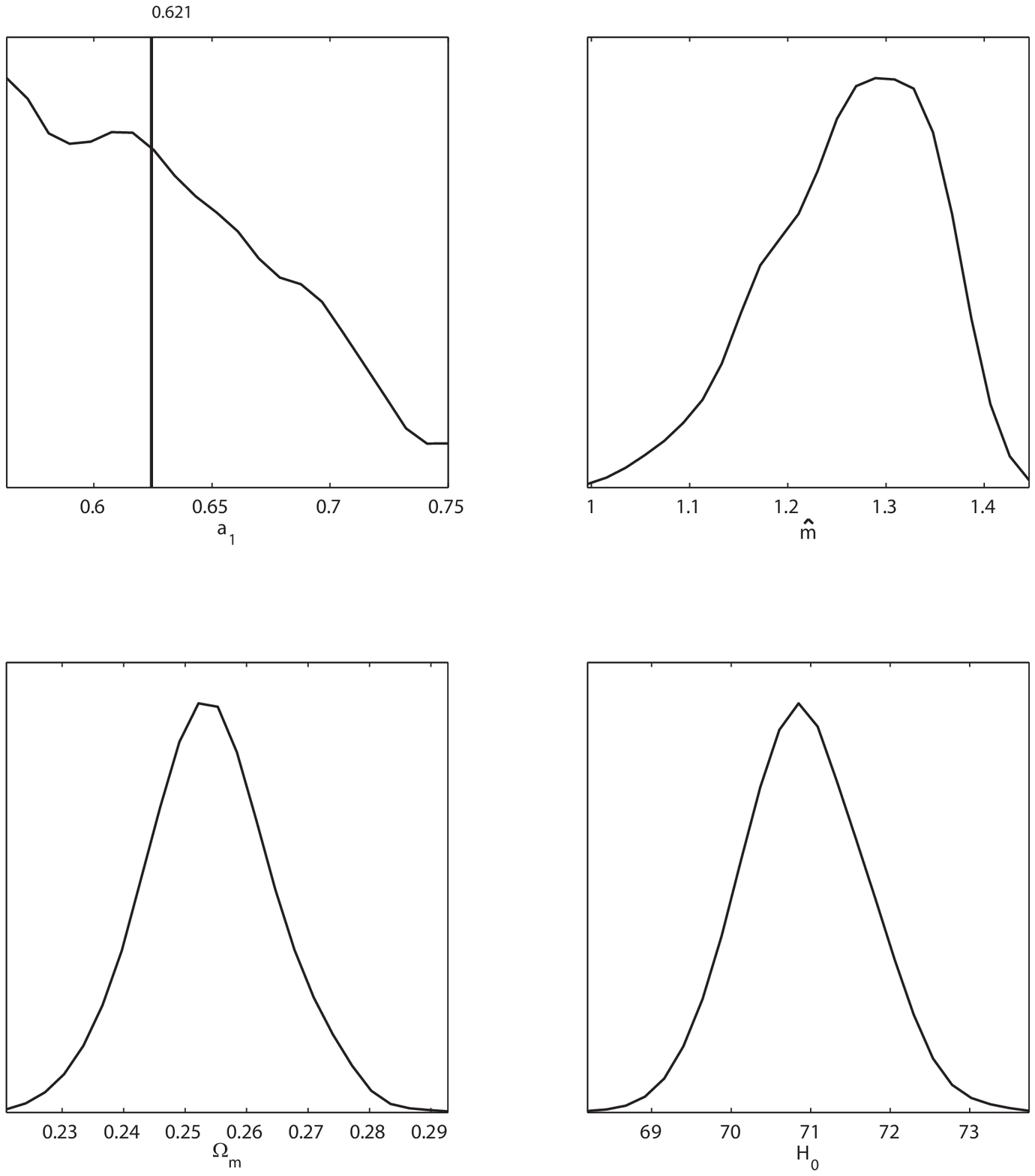}} \\
\hline
\end{tabular}
\caption{
LEFT: HOG models with $n=1$. $2D$ joint contour plots for (clockwise from top left) ($a_1$,$\Omega_m$),  ($\hat{m}$,$\Omega_m$), ($\hat{m}$,$H_0$), and ($H_0$,$\Omega_m$) planes for WMAP5, SDSS LRG(BAO), Union SNe Ia data sets, Hubble Key Project, and age of the universe where the inner and outer loops are $68\%$ and $95\%$, respectively. On the top left, the region to the left of the vertical bar is the one satisfying the physical conditions of table I. RIGHT: 1D parameter distributions for the same data set as at LEFT. The vertical line represents the cut for physical conditions in Table 1.} 
\end{center}
\end{figure}

Following \cite{Komatsu2008}, we define three fitting parameters for comparison to WMAP5 data, the shift parameter, $R$, 
\be
R(z_*)=\sqrt{\Omega_m}H_0(1+z_*)D_A(z_*),
\label{eq:ShiftParameter}
\ee
with the redshift, $z_*$ for the surface of last scattering, 
\be
z_*=1048[1+0.00124(\Omega_b h^2)^{-0.738}][1+g_1 (\Omega_m h^2)^{g_2}],
\label{eq:zstar}
\ee
where 
\be
g_1=\frac{0.0783(\Omega_b h^2)^{-0.238}}{1+39.5(\Omega_b h^2)^{0.763}},
\label{eq:zstarg1}
\ee
and
\be
g_2=\frac{0.560}{1+21.1(\Omega_b h^2)^{1.81}},
\label{eq:ztarg2}
\ee
and third, the acoustic scale, $l_a$, is
\be
l_a=(1+z_*)\frac{\pi D_A(z_*)}{r_s(z_*)},
\label{eq:AcousticScale}
\ee
with the proper angular diameter distance, $D_A(z)=D_L(z)/(1+z)^2$ and the comoving sound horizon, $r_s(z_*)$ 
\be
r_s(z_*)=\frac{1}{\sqrt{3}}\int^{1/(1+z_*)}_{0}{\frac{da}{a^2H(a)\sqrt{1+(3\Omega_b/4\Omega_{\gamma})a}}},
\label{eq:SoundHorizon}
\ee
with $\Omega_{\gamma}=2.469\times 10^{-5}h^{-2}$ for $T_{cmb}=2.725 K$.

Together the parameters $x_i=(R,l_a,z_*)$ are used to fit $\chi^2_{CMB}=\triangle x_iCov^{-1}(x_ix_j)\triangle x_j$ with $\triangle x_i=x_i-x^{obs}_i$ and $Cov^{-1}(x_ix_j)$is the inverse covariance matrix for the parameters. 

Next, for the BAO, we follow \cite{PercivalEisensteinHuTegmark} and define the ratio of the sound horizon, $r_s(z_d)$ to the effective distance, $D_V$ as a fit for SDSS by
\be
\chi^2_{BAO}=\Big(\frac{r_s(z_d)/D_V(z=0.2)-0.198}{0.0058}\Big)^2+\Big(\frac{r_s(z_d)/D_V(z=0.35)-0.1094}{0.0033}\Big)^2,
\label{eq:chisqbao}
\ee
with 
\be
D_V(z)=\Big(D_A^2(z)(1+z)^2\frac{z}{H(z)}\Big)^{1/3},
\label{eq:EffectiveDistance}
\ee
and the redshift, $z_d$ as
\be
z_d=\frac{1291(\Omega_m h^2)^{0.251}}{1+0.659(\Omega_m h^2)^{0.828}}[1+b_1(\Omega_b h^2)^{b_2}],
\label{eq:zdrag}
\ee
where
\be
b_1=0.313(\Omega_m h^2)^{-0.419}[1+0.607(\Omega_m h^2)^{0.674}],
\label{eq:zdragb1}
\ee
and
\be
b_2=0.238(\Omega_m h^2)^{0.223}.
\label{eq:zdragb2}
\ee

Our results for n=1 are presented in Fig. 5 showing how the combination of SN+CMB{\_}distance+BAO, along with Hubble Key Project and the age of the universe prior, provides tighter constraints on the HOG models considered. We find the best-fit $\chi_{SN+BAO+CMB}^2\approx 316.8$ compared to $\chi_{SN+BAO+CMB}^2\approx314.6$ for the LCDM concordance model. We find $\Omega_m=0.252^{+0.024}_{-0.021}$ and $\hat{m}=1.338^{+0.085}_{-0.232}$. For the parameter $a_1$, we obtain now better observational constraints with $a_1=0.569^{+0.052}_{-0.006}$, but as shown on the Fig. 5a, the cuts there are still coming from the physical acceptability condition $0.563\le a_1\le 0.621$ (see Table 1). We also get tighter constraints on $H_0=71.00^{+1.67}_{-1.70}$. Next, for $n=2$ models, we obtain a $\chi_{SN+BAO+CMB}^2\approx 316.2$ and the model parameters are also better constrained with the additional data. The best fit values are $\Omega_m=0.253^{+0.023}_{-0.018}$ and $\hat{m}=1.284^{+0.134}_{-0.214}$. Also, the constraints on the model parameter $a_1$ are now tighter, $a_1=0.692^{+0.041}_{-0.009}$, but again the cuts come from the physical acceptability condition $0.683 \le a_1\le 0.734$. We also get tighter constraints on $H_0=70.55^{+1.59}_{-1.21}$. Our results for $n=2$ are in Fig. 6. Finally, we find again that because of the possible systematic uncertainties in the data considered, it is not clear that the difference between the $\chi^2$ achieved by these models and the one from the LCDM model is significant.
\begin{figure}
\begin{center}
\begin{tabular}{|c|c|}
\hline
{\includegraphics[width=2.8in,height=2.8in,angle=0]{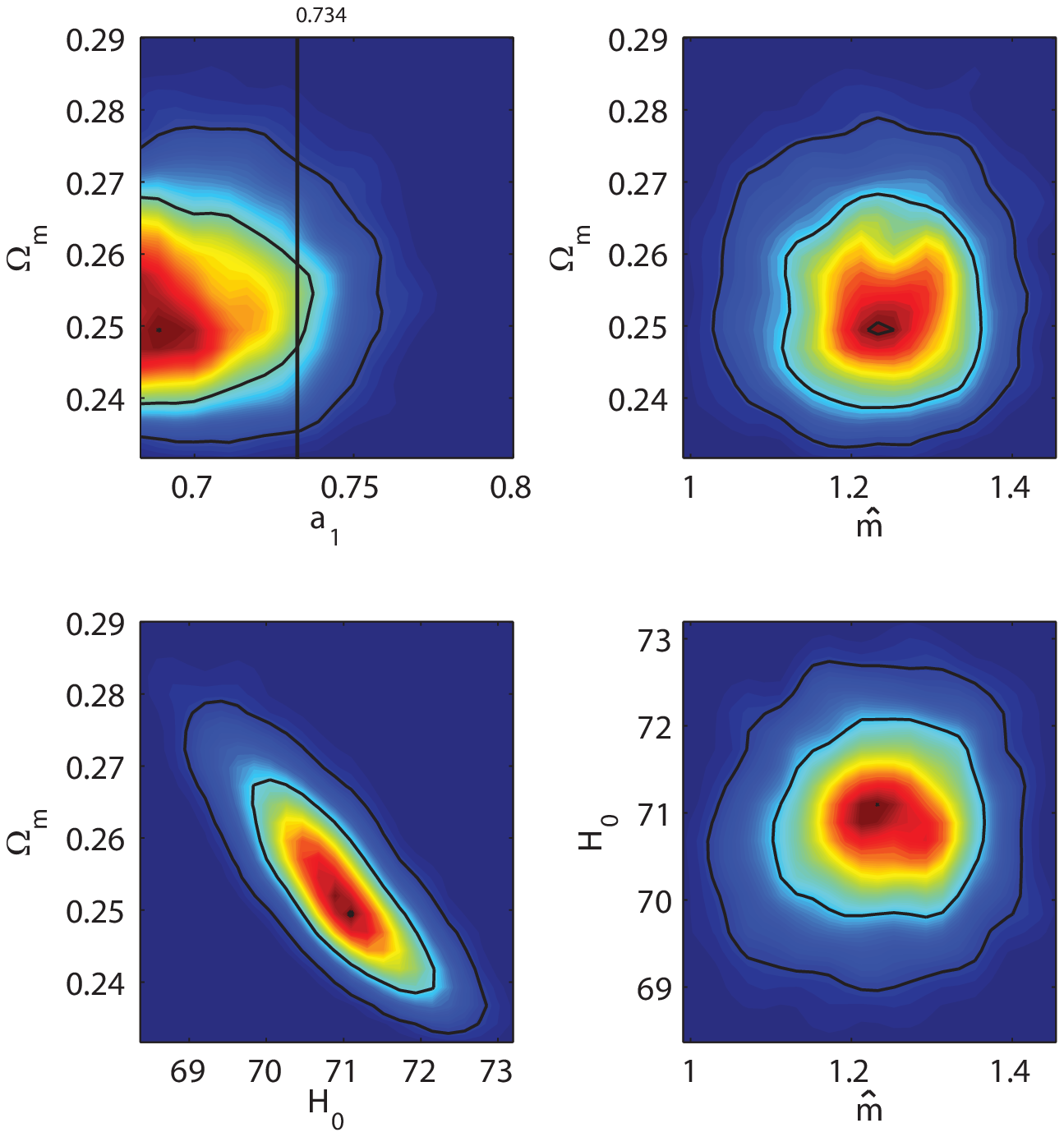}} &
{\includegraphics[width=2.8in,height=2.8in,angle=0]{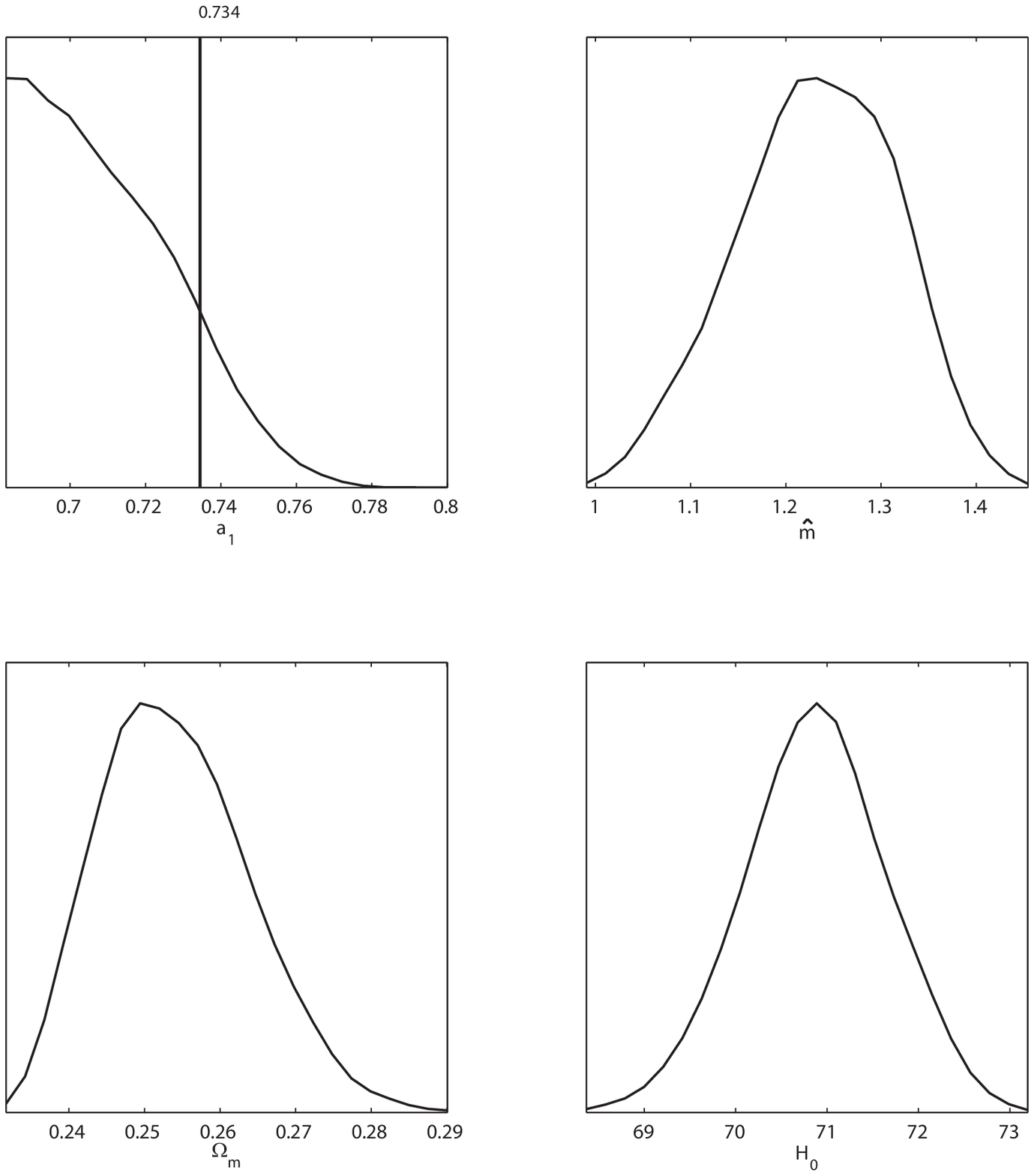}} \\
\hline
\end{tabular}
\caption{
LEFT: HOG models with $n=2$. $2D$ joint contour plots for (clockwise from top left) ($a_1$,$\Omega_m$),  ($\hat{m}$,$\Omega_m$), ($\hat{m}$,$H_0$), and ($H_0$,$\Omega_m$) planes for WMAP5, SDSS LRG(BAO), Union SNe Ia data sets, Hubble Key Project, and age of the universe where the inner and outer loops are $68\%$ and $95\%$, respectively. On the top left, the region to the left of the vertical bar is the one satisfying the physical conditions of table I.   RIGHT: 1D parameter distributions for the same data set as at LEFT. The vertical line represents the cut for physical conditions in Table 1.} 
\end{center}
\end{figure}

\section{Conclusion}

Following some initial work in \cite{IshakMoldenhauer2008}, we confront here some of the higher order gravity models (HOG), as expressed in terms of a minimal set of curvature invariants, to some physical acceptability conditions and observational constraints. 
The models studied had previously been constrained in \cite{IshakMoldenhauer2008} by the requirement of a self-acceleration phase at late times, a deceleration phase during matter domination, and the avoidance of a separatrix region which would prevent a transition between the two evolutionary phases of the universe. Here, we add the further physical acceptability conditions for the models to be free of ghost instabilities and to have real and positive speeds of mode propagations. When these conditions are imposed, the allowable parameter space for the models considered is significantly reduced. We do find models that meet all these conditions during both accelerating and decelerating phases, but, the parameter space range for which all the modes propagate subluminally is found in conflict with either the separatrix-free condition or the ghost-free ranges. 
We find that if we impose on the models to satisfy all the conditions above during deceleration and acceleration phases, then they will have modes with superluminal propagations during one phase or the other. 
We compare these models to supernovae Ia, distance to the CMB surface, and BAO cosmological constraints. We find a parameter space for the models considered that fit well the data and provide comparable fits to those achieved by the LCDM concordance model. The numerical integration over the full redshift range is elaborate for these models and required the use of methods to deal with stiff non-linear ODEs. We found that the combination of an approximate solution at high redshifts to a full numerical integration at low redshifts ($z<5$) provides a stable and precise integration scheme to compare these and other HOG models to observations. 
While we find HOG models that pass the physical acceptability conditions above and cosmological constraints from supernovae, distance to CMB surface, and BAO constraints, it seems that the limitation of the models studied here comes from the presence of superluminal mode propagations. 
\acknowledgements 
The authors thank D. Easson, K. C. Chan, Y. S. Song, J. Weller, and W. Rindler for useful comments. The authors thank B. Troup and J. Scott for useful discussions about the CosmoMC package. This material is based upon work supported in part by NASA under grant NNX09AJ55G. Part of the calculations for this work have been performed on the Cosmology Computer Cluster funded by the Hoblitzelle Foundation.


\begin{thebibliography}{}
%
\bibitem{observations} 

A. G. Riess, {\em{et al.}}, Astron. J. {\textbf{116}}, 1009-1038 (1998);
%
 S. Perlmutter, {\em{et al.}}, Astrophys. J. {\textbf{517}}, 565-586 (1999);
%
R. A. Knop, {\em{et al.}}, Astrophys. J. {\textbf{598}}, 102-137
(2003); 
%
A. G. Riess, {\em{et al.}}, Astrophys. J. {\textbf{607}}, 665-687
(2004);  
%
C. L. Bennett, {\em{et al.}},
Astrophys. J. Suppl. Ser. {\textbf{148}}, 1 (2003); 
D. N. Spergel, {\em{et al.}},
Astrophys. J. Suppl. Ser. , 175 (2003);
L. Page et al., 
Astrophys. J. Suppl. Ser. {\textbf{148}}, 2333 (2003).
%
 Seljak et al., Phys.Rev. D{\textbf{71}}, 103515 (2005);
%
M. Tegmark, {\em{et al.}},  Astrophys. J. {\textbf{606}}, 702-740
(2004).
%
D.N. Spergel , {\em{et al.}}, 
Astrophys.J.Suppl. 170, 377 (2007).

\bibitem{reviews} S. Weinberg
Rev. Mod. Phys., 
\textbf{61}, 1 (1989);
%
M.S. Turner, 
\physrep, 
\textbf{333}, 619 (2000);
%
V. Sahni, A. Starobinsky
Int.J.Mod.Phys. 
\textbf{D9}, 373 (2000);
%
S.M. Carroll, 
Living Reviews in Relativity, 
\textbf{4}, 1 (2001);
%
T. Padmanabhan, 
\physrep, 
\textbf{380}, 235 (2003);
%
P.J.E. Peebles and B. Ratra, 
Rev. Mod. Phys. 
\textbf{75}, 559 (2003);
%
A. Upadhye, M. Ishak, P. J. Steinhardt, 
Phys. Rev. D
\textbf{72}, 063501 (2005);
%
A. Albrecht et al, {\it Report of the Dark Energy Task Force}
astro-ph/0609591 (2006).
%
M. Ishak, 
Foundations of Physics Journal, 
Vol.\textbf{37}, No 10, 1470 (2007).  
%
\bibitem{Biblio300}We conducted a thorough bibliography search on NASA ADS system, HEP-SPIRES and arXiv:[astro-ph][gr-qc] archives.
%
\bibitem{IshakMoldenhauer2008} M. Ishak and J. Moldenhauer, JCAP 0901:024 (2009).
%

\bibitem{Carroll2005} S. M. Carroll, A. De Felice, V. Duvvuri, D. A. Easson, M. Trodden, M. S. Turner, Phys. Rev. D \textbf{71} 063513 (2005).
%
\bibitem{HOGgroupJAM}   D.A. Easson, Int. J. Mod. Phys. A19, 5343 (2004);
%
F. S. N. Lobo, arXiv:0807.1640v1 [gr-qc] (2008).
%
V. Faraoni, Phys. Rev. D \textbf{74} 023529 (2006).
%
A.D. Dolgov, M. Kawasaki, Phys. Lett. B \textbf{573} 1 (2003).
%
L. Amendola, R. Gannouji, D. Polarski, S. Tsujikawa, Phys. Rev. D \textbf{75} (2007) 083504.
%
J. Moldenhauer, M. Ishak, in preparation, 2009.
%
G. Cognola, E. Elizalde, S. Nojiri, S. D. Odintsov, L. Sebastiani, S. Zerbini, arxiv:0712.4017.
%
I. Brevik, J. Q. Hurtado, arxiv:gr-qc/0610044.
%
T. P. Sotiriou, V. Faraoni, arxiv:0805.1726.
%
R. P. Woodward, Lect. Notes Phys. \textbf{720} 403 (2007).
%
S. Nojiri, S. D. Odintsov, Phys. Lett. B \textbf{631} 1 (2005).
%
V. Faraoni, Presented at SIGRAV2008, 18th Congress of the Italian Society of General Relativity and Gravitation, Cosenza, Italy September 22-25, 2008, arxiv:0810.2602.
%
K.~i.~Maeda and N.~Ohta,  Phys.\ Lett.\  B {\bf 597} (2004) 400;
%
K.~i.~Maeda and N.~Ohta, Phys.\ Rev.\  D {\bf 71} (2005) 063520;
%
K.~Akune, K.~i.~Maeda and N.~Ohta, Phys.\ Rev.\  D {\bf 73} (2006) 103506;
%
N.~Ohta, Int.\ J.\ Mod.\ Phys.\  A {\bf 20} (2005).
%
\bibitem{Faraoni2006b} V. Faraoni, Phys. Rev. D \textbf{74} 104017 (2006).
%
\bibitem{Sotiriou2007} T. P. Sotiriou, Ph. D. Thesis, arxiv:0710.4438 (2007).
%
\bibitem{Sotiriou2006} T. P. Sotiriou, Class. Quant. Grav. \textbf{23},  1253 (2006).
%
\bibitem{MengWang2004} X. Meng, P. Wang, Class. Quant. Grav. \textbf{21}, 2029 (2004).
%
\bibitem{Stelle1977} K. S. Stelle, Phys. Rev. D \textbf{16} 953 (1977).
%
N. H. Barth and S. M. Christensen, Phys. Rev. D \textbf{28} 1876 (1983).
%
\bibitem{BirrellDavies1982} N.D. Birrell and P.C.W. Davies, \textit{Quantum Fields in Curved Space} (Cambridge University Press 1982).
%
\bibitem{Mena2006} O. Mena, J. Santiago, J. Weller, Phys. Rev. Lett. \textbf{96}, 041103 (2004).
%
\bibitem{Shirata2005} A. Shirata, T. Shiromizu, N. Yoshida, Y. Suto, Phys. Rev. D \textbf{71} 064030 (2005).
%
\bibitem{Borowiec2006} A. Borowiec, W. Godlowski, M. Szydloski, Phys. Rev. D \textbf{74} 043502 (2006).
%
\bibitem{NavarroVanAcoleyen2005} I. Navarro, K. Van Acoleyen, Phys. Lett. B \textbf{622} 1 (2005).
%
\bibitem{Chiba2003} T. Chiba, Phys. Lett. B \textbf{575} 1 (2003);
%
T. Chiba, T. L. Smith, A. L. Erickcek, Phys. Rev. D \textbf{75} 124014 (2007).
%
\bibitem{Debever1964} R. Debever, Cah. Phys. \textbf{168} 303 (1964).
%
\bibitem{CarminatiMcLenaghan1991}  J. Carminati, R. McLenaghan, J. Math. Phys. \textbf{32}(11) 3135 (1991).
%
\bibitem{ZakharyMcIntosh1997} E. Zakhary C. McIntosh, Gen. Rel. Grav. \textbf{29} (5) 539 (1997).


\bibitem{Segre}
C. Segre, {\textit{Memorie della R. Accademia dei Lincei, ser. 3a XIX, 127, Sec 5.1}} (1884). 
%
%
\bibitem{Thomas1934} T. Y. Thomas, \textit{The Differential Invariants of Generalized Spaces} (Cambridge University Press 1934).
%
V. V. Narlikar, K. R. Karmarkar, Proc. Ind. Acad. Sci. A \textbf{29} 91 (1948).
%
J. Geheniau, R. Debever, Bull. Cl. Sci. Acad. R. Belg. \textbf{XLII} 114 (1956).
%
R. Debever, Bull. Cl. Sci. Acad. R. Belg. \textbf{XLII} 252 (1956).
%
L. Witten, Phys. Rev. \textbf{113} 357 (1959).
%
A. Z. Petrov, \textit{Einstein Spaces} (Pergamon, Oxford 1969).
%
G. L. Santosuosso, \textbf{35} (7) 1307 (1999).  
%
J. Carminati, E. Zakhary, Class. Quant. Grav. \textbf{16} 3221 (1999).
%
\bibitem{Stephani2003} H. Stephani, D. Kramer, M. MacCallum, C. Hoenselaers, E. Herlt \textit{Exact Solutions of Einstein's Field Equations} (2nd edn.) (Cambridge University Press 2003).
%
\bibitem{Petrov1954} A. Z. Petrov, Uch. Zapiski Kazan Gos. Univ. \textbf{144} (1954);  Reprint: Gen. Rel. Grav. \textbf{32} 16665 (2000).
%
\bibitem{Pirani1957} F. E. A. Pirani, Phys. Rev. \textbf{105} 1089 (1957).
%
\bibitem{Penrose1960} R. Penrose, Ann. Phys. \textbf{10} 171 (1960). 
%
\bibitem{Chiba2005} T. Chiba, JCAP \textbf{0503} 008 (2005).
%
\bibitem{Dvali2006} G. Dvali, New J.Phys. \textbf{8} 326 (2006).
%
\bibitem{LiBarrowMota2007} B. Li, J. D. Barrow, D. F. Mota, Phys. Rev. D \textbf{76} 044027 (2007).
%
\bibitem{DeFelice2006} A. De Felice, M. Hindmarsh, M. Trodden, JCAP \textbf{0608} 005 (2006).
%
\bibitem{Calcagni2006} G. Calcagni, B. de Carlos, A. De Felice, Nucl.Phys. B752  404-438 (2006).
%
\bibitem{NavarroVanAcoleyen2006} I. Navarro, K. Van Acoleyen, JCAP 0603 008 (2006).
%
\bibitem{DeFelice2006b} A. De Felice, M. Hindmarsh, JCAP \textbf{0706} 028 (2007).
%
\bibitem{GoheerDunsby2009} N. Goheer, R. Goswami, P. K. S. Dunsby, K. Ananda arxiv:0904.2559.
%
\bibitem{Uddin2009} K. Uddin, J. E. Lidsey, R. Tavakol, arxiv:0903.0270.
%
\bibitem{ZhouCopeland2009} S. Y. Zhou, E. J. Copeland, P. M. Saffin, arxiv:0903.4610.
%
\bibitem{DeFelice2008} A. De Felice, S. Tsujikawa, arxiv:0810.5712.
%
\bibitem{HwangNoh2000} J. C. Hwang and H. Noh, Phys. Rev. D \textbf{61} 043511 (2000).
%
\bibitem{HwangNoh2005} J. C. Hwang and H. Noh, Phys. Rev. D \textbf{71} 063536 (2005).
%
\bibitem{KoivistoMota2007a} T. Koivisto and D. F. Mota, Phys.Rev. D75 023518 (2007).
%
\bibitem{KoivistoMota2007b} T. Koivisto and D. F. Mota, Phys.Lett. B644 104-108 (2007).
%
\bibitem{AmendolaDavis2006} L. Amendola, C. Charmousis, S. C. Davis, 
JCAP \textbf{0612} 020 (2006).
%
\bibitem{Cognola2006} G. Cognola, E. Elizalde, S. Nojiri, S. D. Odintsov, S. Zerbini, Phys. Rev. D \textbf{73} 084007 (2006).
%
\bibitem{NojiriOdintsov2005} S. Nojiri, S. D. Odintsov, Phys. Lett. B \textbf{631} 1 (2005).
%
\bibitem{NojiriOdintsov2008} S. Nojiri, S. D. Odintsov, arxiv:0801.4843.
%
\bibitem{NojiriOdintsov2006} S. Nojiri, S. D. Odintsov, Int.J.Geom.Meth.Mod.Phys. 4 115-146 (2007).  
%
\bibitem{Nojiri2005} S. Nojiri, S. D. Odintsov, M. Sasaki, Phys. Rev. D \textbf{71} (2005) 123509.
%
\bibitem{Kowalski2008} M. Kowalski, et. al. Astrophys.J.686:749-778, (2008); Union data sets include
%
Hamuy et. al. , AJ, 112, 2408 (1996);
%
Krisciunas et. al. , AJ, 127, 1664 (2004a),  AJ, 128, 3034 (2004b), AJ, 122, 1616 (2001);
%
Riess et. al., AJ, 116, 1009 (1998), AJ, 117, 707 (1999), ApJ, 607, 665 (2004),  ApJ, 659, 98 (2007);
%
Jha et. al. , AJ, 131, 527 (2006), ApJ, 659, 122 (2007);
%
Perlmutter et. al. , ApJ, 517, 565 (1999);
%
Tonry et. al., ApJ, 594, 1 (2003);
%
Barris et. al. , ApJ, 602, 571, (2004);
%
Knop et. al. , ApJ, 598, 102 (2003);
%
Astier et. al.,  A and A, 447, 31, (2006);
%
Miknaitis et. al., ApJ, 666, 674 (2007);
%
Wood-Vasey et. al., ApJ, 666, 694 (2007);
%
Garnavich et. al., ApJ, 509, 74 (1998);
%
Schmidt et. al., ApJ, 507, 46 (1998).
%
\bibitem{LewisBridle2001} A. Lewis and S. Bridle, Phys. Rev. D \textbf{66} 103511 (2002).
%
\bibitem{SongPeirisHu2006} Y. S. Song, H. Peiris, W. Hu, Phys. Rev. D \textbf{76} 063517 (2007).
%
\bibitem{LiChanChu2006} B. Li, K. C. Chan,  M. C. Chu, Phys.Rev.D76:024002 (2007).
%
\bibitem{HubbleKeyProject} W. L. Freedman et. al. Astrophys. J \textbf{553}, 47 (2001).
%
\bibitem{KraussChaboyer2003} L. M. Krauss and B. Chaboyer, Science \textbf{299}, 65 (2003).
%
\bibitem{Komatsu2008} Komatsu et. al. Astrophys.J.Suppl.180:330-376 (2009).
%
\bibitem{PercivalEisensteinHuTegmark} W. Percival, et. al. Mon.Not.Roy.Astron.Soc.381:1053-1066 (2007);
%
D. J. Eisenstein et. al.,  ApJ, 633, 560 (2005);
%
W. Hu and N. Sugiyama, Astrophys.J. 471 (1996) 542-570;
%
M. Tegmark et. al. Phys.Rev.D74:123507 (2006).
%
\bibitem{Weller2006} J. Weller, Talk from XLI Recontres de Moriond March, 18-25, (2006).
%
\bibitem{Davis2007} T. M. Davis, accepted APJ, arxiv:astro-ph/0701510, (2007).
%
\end{thebibliography}
\end{document}